\documentclass[mathleft
]{an}
\usepackage{graphicx}
\usepackage{times}
\usepackage{epsfig}
\begin{document}

\Pagespan{1}{6}
\Yearpublication{2015}%
\Yearsubmission{2015}%
\Month{11}%
\Volume{999}%
\Issue{88}%

\title{Unwrapping the X-ray Spectra of Active Galactic Nuclei}

\author{Christopher S. Reynolds\inst{1,2}\fnmsep\thanks{Corresponding author:
  \email{chris@astro.umd.edu}\newline}}
\titlerunning{Unwrapping the X-ray Spectra of AGN}
\authorrunning{Christopher S. Reynolds}
\institute{
Department of Astronomy, University of Maryland, College Park, MD~20742, USA
\and 
Joint Space Science Institute, University of Maryland, College Park, MD~20742, USA
}

\received{1 Sep 2015}
\accepted{-}
\publonline{-}


\abstract{Active galactic nuclei (AGN) are complex phenomena.  At the heart of an AGN is a relativistic accretion disk around a spinning supermassive black hole (SMBH) with an X-ray emitting corona and, sometimes, a relativistic jet. On larger scales, the outer accretion disk and molecular torus act as the reservoirs of gas for the continuing AGN activity. And on all scales from the black hole outwards, powerful winds are seen that probably affect the evolution of the host galaxy as well as regulate the feeding of the AGN itself.  In this review article, we discuss how X-ray spectroscopy can be used to study each of these components.  We highlight how recent measurements of the high-energy cutoff in the X-ray continuum by NuSTAR are pushing us to conclude that X-ray coronae are radiatively-compact and have electron temperatures regulated by electron-positron pair production.  We show that the predominance of rapidly-rotating objects in current surveys of SMBH spin is entirely unsurprising once one accounts for the observational selection bias resulting from the spin-dependence of the radiative efficiency.  We review recent progress in our understanding of fast ($v\sim 0.1-0.3c$), highly-ionized (mainly visible in FeXXV and FeXXVI lines), high-column density winds that may dominate quasar-mode galactic feedback.  Finally, we end with a brief look forward to the promise of Astro-H and future X-ray spectropolarimeters. }

\maketitle

\section{Introduction}

The accretion of matter onto a black hole is a remarkably complex process.  On scales of the black hole itself, the dynamics of the accretion flow and the spectrum/variability of the emitted radiation is strongly influenced by general relativistic (GR) effects.   The interaction of large-scale magnetic fields with the spinning event horizon of the black hole can readily create relativistic jets (Blandford \& Znajek 1977) which may themselves be an important source of observed radiation.  In the central regions of the accretion disk, a significant fraction of the accretion energy is channeled into a hot ($\sim10^9$\,K), optically-thin, X-ray emitting corona (Haardt \& Maraschi 1991) which, in non-blazar sources, is thought to dominate the observed X-ray emission.  On larger scales, powerful winds can be driven from the accretion disk via radiative, magnetic, or thermal pressure forces.  For the accreting supermassive black holes (SMBH) that are the focus of this paper, these winds can strongly shape the galactic environment in which they reside, stifling star formation and modulating the fueling of the active galactic nucleus (AGN).   

\begin{figure*}
\centerline{
\psfig{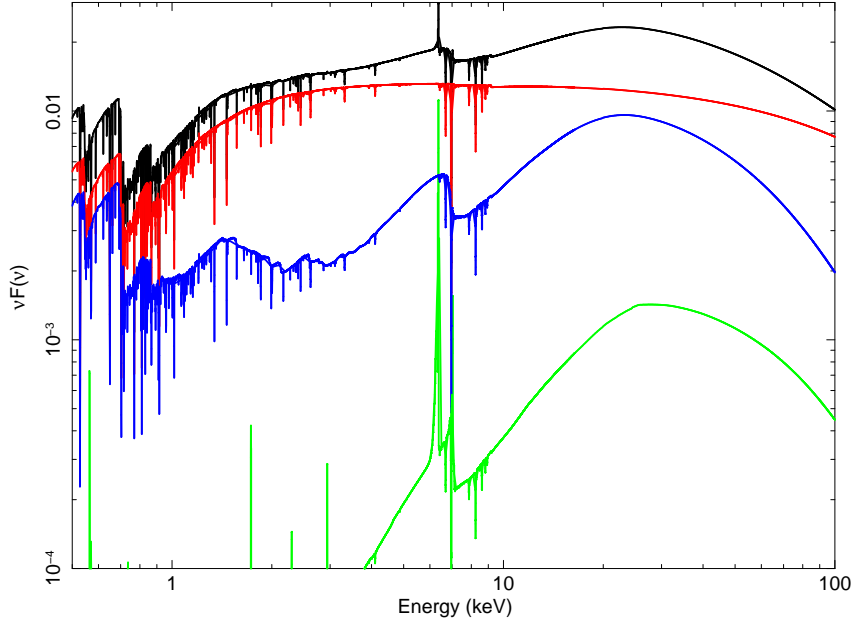}}
\caption{The basic X-ray spectral components of a (non-blazar) AGN, illustrated with a model based on the Chiang \& Fabian (2011) study of the Seyfert galaxy MCG--6-30-15.  The primary X-ray continuum emitted by the corona (shown in red) has the form of a power law with photon index $\Gamma\sim 2$ and a high-energy rollover at $E \sim {\rm few}\times 100$\,keV.   This continuum irradiates any optically-thick matter in the vicinity, creating so-called X-ray reflection signatures (Basko 1978; George \& Fabian 1991; models shown here due to Garcia et al. 2014).  Reflection from any cold, distant, slowing moving material such as the molecular torus of AGN unification schemes (shown in green) mainly shows itself as a narrow 6.4\,keV iron fluorescence line.  Reflection from the ionized inner accretion disk (shown in blue) produces a strongly broadened and redshifted iron-K$\alpha$ line (Fabian et al. 1989) as well as soft and hard excesses (due to radiative-combination and Compton scattering respectively).  All of these emission components are then subject to absorption by a surrounding accretion disk wind that can imprint a rich spectrum of absorption lines and bound-free edges (Reynolds 1997). }
\end{figure*}

In this paper, we review how the detailed analysis of current and future high-quality X-ray spectra of AGN allow us to unwrap and study much of this complexity. For the reader who may be unfamiliar with the basic X-ray spectral components of (non-blazar) AGN, these are described in Fig.~1 and its associated caption.   We start (Section~2) with a discussion of the physics of the X-ray corona itself, including recent progress that has been enabled by hard X-ray measurements of the continuum cutoff energy.  We then summarize the status of SMBH spin studies, and show that the predominance of rapidly spinning black holes in current samples is entirely expected given reasonable spin-dependent selection effects (Section~3).  Section~4 describes recent progress in the study of the accretion disk winds that may be relevant for galactic scale feedback.   Finally, we discuss the future outlook in Section~5.

\section{The physics of the X-ray emitting corona}

While the accretion disk in a luminous AGN is expected to emit thermal UV radiation, the  observed X-ray emission is attributed to thermal Comptonization in a hot and optically-thin accretion disk corona.  However, many aspects of the corona's physical nature are still very uncertain, including its geometry (slab, sphere, patches, or jet?), the energy source (SMBH spin or accretion energy?), and the role of thermal versus non-thermal processes.  

NuSTAR (Harrison et al. 2013) has provided a major advance by permitting the measurements of the high-energy continuum cutoff, and hence the (electron) temperature of the coronal plasma, in a sample of AGN (Fabian et al. 2015).  For cutoff energies of $\sim 150$\,keV or less (corresponding to coronal electron temperatures of $kT_e\sim 50$\,keV or less), the quasi-exponential rollover of the continuum can be directly tracked by NuSTAR (Brenneman et al. 2014).  However, counterintuitively, NuSTAR (which has an operating bandpass of 3--80\,keV) can provide meaningful constraints on much higher cutoff energies provided that the source has a moderately strong disk reflection spectrum (Garcia et al. 2015) --- this is because the temperature structure of the outermost layers of the accretion disk atmosphere, and hence the detailed form of the reflection spectrum, is sensitive to the high-energy cutoff (left panels of Figure~2).

\begin{figure*}[t]
\hbox{
\psfig{figure=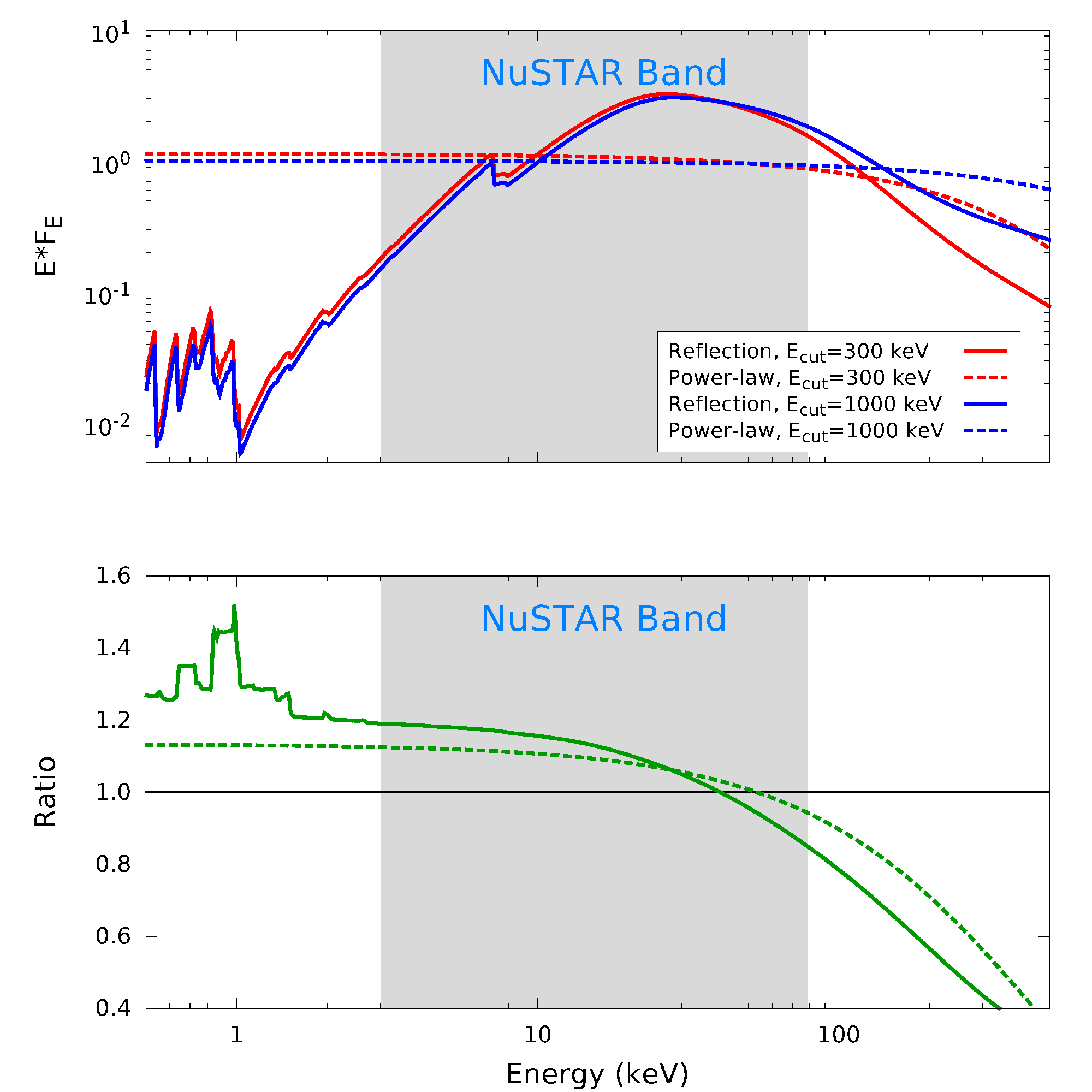,width=0.44\textwidth}
\psfig{figure=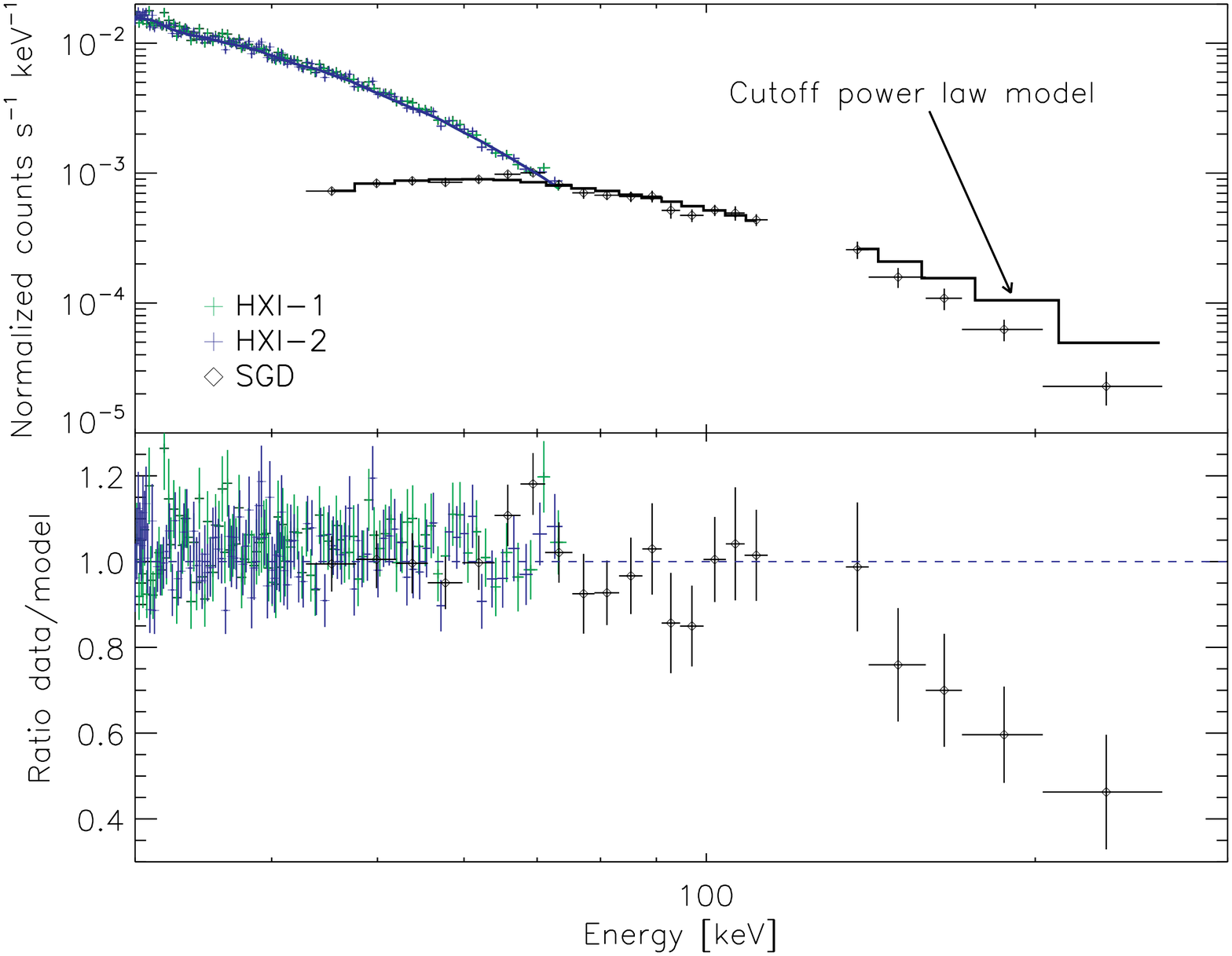,width=0.54\textwidth}
}
\caption{{\it Left Panels  : }Demonstration of the sensitivity of the accretion disk reflection spectrum to the high-energy cutoff $E_{\rm cut}$.  Even for a continuum cutoff that is well outside of the NuSTAR bandpass, it imprints itself (via its effect on the temperature structure of the disk atmosphere) on both the Compton reflection hump and the soft X-ray emission lines.  Figure reproduced from Garcia et al. (2015).   {\it Right Panel : }Simulation of a 100\,ks Astro-H Hard X-ray Imager (HXI) and Soft Gamma-Ray Detector (SGD) spectrum of the bright Seyfert galaxy NGC4388.  The spectral model used for the simulated data employs the {\tt compPS} physical Comptonization model (Poutanen \& Svensson 1996), using $kT_e=50$\,keV and $\tau_e=1$.  Displayed are the ratios when the simulated data are then fitted with a simple (exponential) cutoff power law model; the fit is clearly poor, demonstrating that the data are capable of characterizing the detailed deviations from the exponential cutoff model.  Figure reproduced from the Astro-H White Paper ``AGN Reflection'' by Reynolds et al. (2014a). }
\end{figure*}

Samples of AGN continuum cutoff energies from NuSTAR have already revealed an important aspect of coronal physics.  Thirty years ago, it was pointed out that for a given coronal compactness $\ell$, defined in dimensionless form as
\begin{equation}
\ell=\frac{L}{R}\frac{\sigma_T}{m_ec^3},
\end{equation}
where $L$ is the luminosity and $R$ is the radius of the corona (Guilbert, Fabian \& Rees 1983), there is a maximum electron temperature set by electron-positron pair production (Svensson 1982, 1984).  Essentially, once the temperature exceeds a threshold, additional energy input goes into pair-production\footnote{In addition, the optical depth of the corona will increase due to the extra pairs, leading to an increased Compton-cooling rate that can more than compensate for the additional heating (Dove, Wilms \& Begelman 1997).} thereby stabilizing the temperature. The result is a ``pair thermostat'' line on the $(\ell, kT_e)$-plane given by
\begin{equation}
\ell\sim 10\left(\frac{kT_e}{m_ec^2}\right)^{5/2}e^{m_ec^2/kT_e}.
\end{equation}
By combining a compilation of NuSTAR-measured coronal temperatures with estimates of coronal compactness from X-ray reverberation studies, Fabian et al. (2015) found that most AGN follow this pair thermostat line, a dramatic confirmation of prior theoretical work and a demonstration that pair production is an important process in real disk coronae.  

With the imminent launch of Astro-H, studies of AGN X-ray coronae will continue their renaissance.  Combining data from all of the instruments onboard Astro-H provides a view of the (instantaneous) 0.5--600\,keV spectrum of a bright AGN, although realistic AGN count rates and background levels may lower the practical upper limit to $<400$\,keV.  With high-quality spectra over such a range, the exponential rollover models that are usually used today to characterize the continuum cutoffs will be wholly inadequate (see right panel of Fig.~2).  Astro-H will be able to determine the precise shape of the continuum rollover, gaining new sensitivity to coronal geometry and the electron distribution function and opening a window on the physics of the electron heating mechanism.

\section{Black hole spin and selection effects}

\begin{figure*}[t]
\hbox{
\psfig{figure=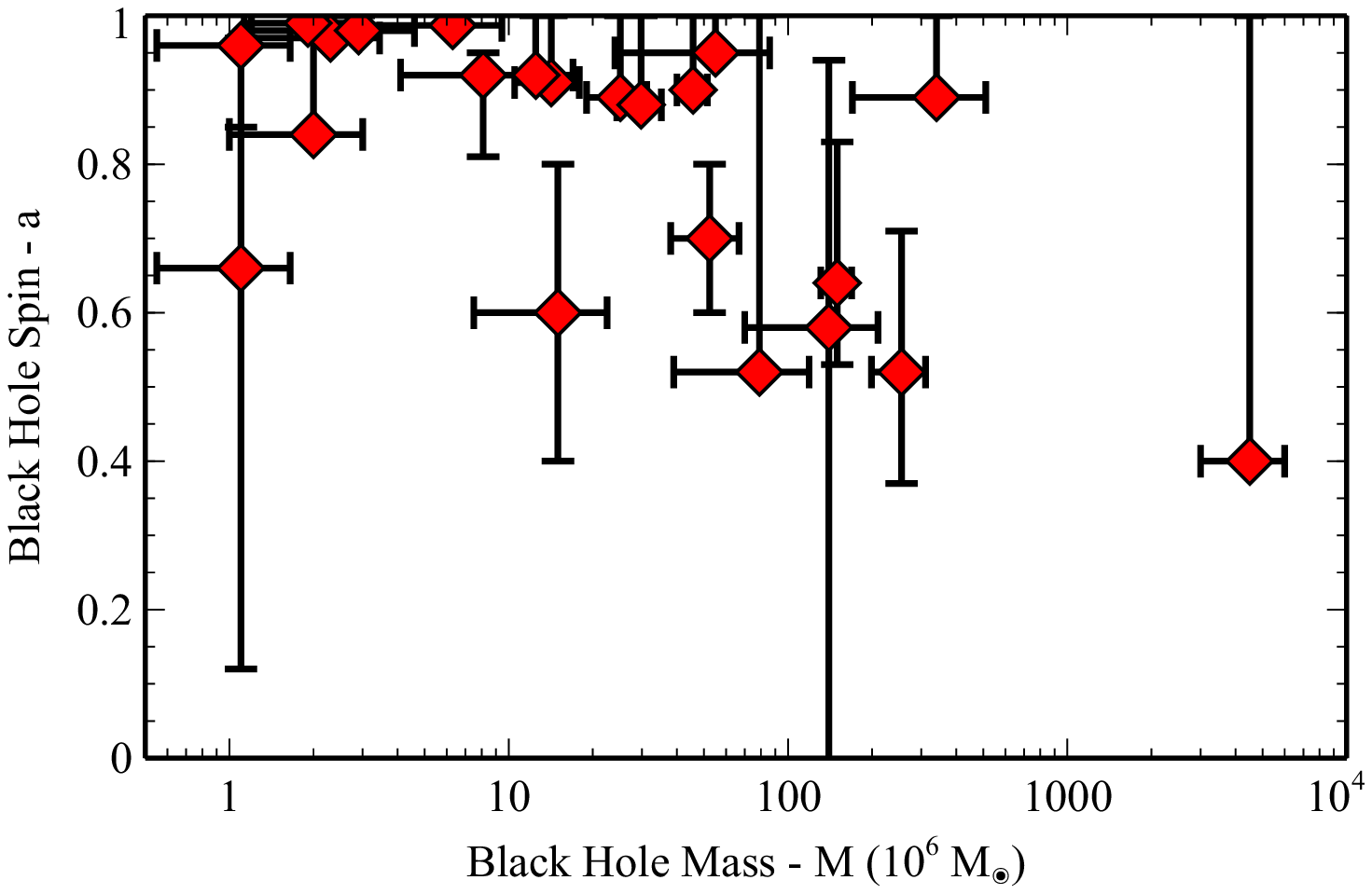,width=0.5\textwidth}
\psfig{figure=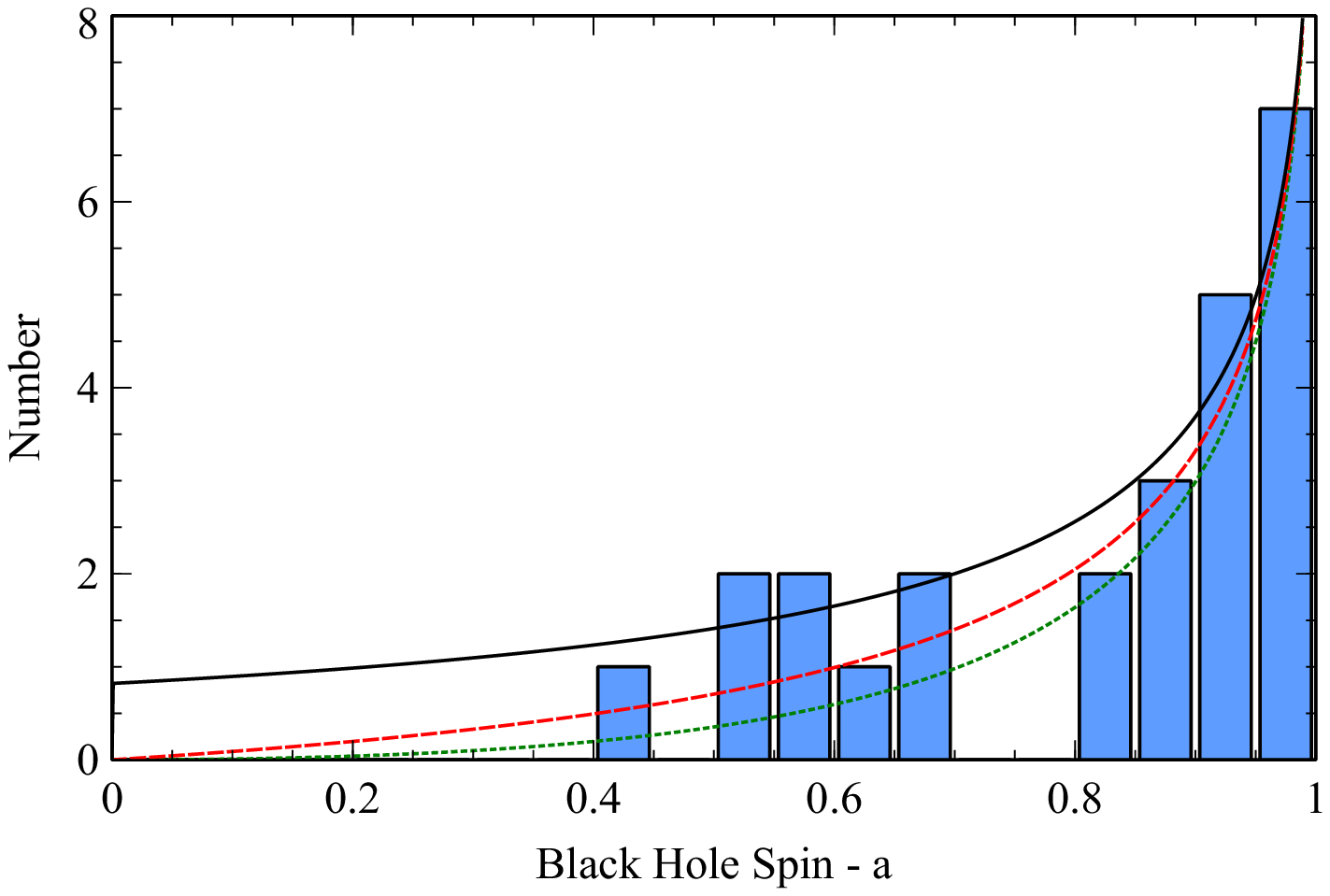,width=0.5\textwidth}
}
\caption{{\it Left Panel : }SMBH mass and spin estimates.  Note that, to conform with conventions from the relevant communities, the masses (mostly derived from optical/UV broad line reverberation measurements) are shown with $1\sigma$-errors whereas the spins (exclusively derived from X-ray reflection) are shown with 90\% errors/limits.  {\it Right Panel : }Histogram of SMBH spins, overlaid with predicted spin distributions that would be observed in a flux-limited sample if the underlying population had $n(a)\propto a^p$ objects per unit spin, where $p=0$ (black solid line), $p=1$ (red dashed line) and $p=2$ (green dotted line).  Note that, in this schematic treatment, objects for which only lower-limits are present on the spin are placed in the bin corresponding to that lower-limit. This is true for the lowest spin value present on the plot, corresponding to the quasar H1821$+$643 for which Suzaku spectroscopy gives $a>0.4$ (Reynolds et al. 2014b).  Figures reproduced from Vasudevan et al. (2015).}
\end{figure*}

The X-ray corona irradiates the inner regions of the SMBH accretion disk and produces the relativistic reflection signatures that have been the focus of a large body of work since the days of ASCA (Tanaka et al. 1995; see reviews by Reynolds \& Nowak 2003, Fabian \& Miniutti 2005, Miller 2007 and references contained within).  A particular emphasis over the past decade has been the use of these signatures for determining the spin of the SMBH (Iwasawa et al. 1996; Dabrowski et al. 1997; Brenneman \& Reynolds 2006).  We defer any detailed discussion of these methods to recent review articles (e.g., Reynolds 2014; Miller \& Miller 2015; Middleton 2015), here simply stating that the X-ray reflection features are thought to truncate close to the GR innermost stable circular orbit (which is a function of spin), and hence have a spin dependent maximum redshift that can be observationally characterized.  

Figure~3 (left panel) shows a compilation of current AGN spin results on the mass-spin plane, as compiled by Vasudevan et al. (2015), updating the similar plot in Reynolds (2014).  It is apparent that there is a large population of rapidly spinning objects amongst this sample.  It is still too early to draw firm conclusions from this fact --- this is an ill-defined sample, consisting of those objects that the community have chosen (for a variety of reasons) to obtain deep X-ray spectra, and the spin parameter of many of the objects have substantial error bars.  It is, however, important to note a powerful selection bias that will strongly promote rapidly spinning black holes into any samples that are even approximately flux-limited.  The more rapidly-spinning black holes have a substantially higher radiative-efficiency and, assuming that the mass accretion rate is set by the astrophysical environment and is uncorrelated with the spin, will be generally more luminous.  To illustrate with an extreme example, standard thin-disk accretion theory predicts a radiative efficiency of $\eta_0\approx 0.06$ for a non-rotating ($a=0$) black hole, but $\eta_{\rm max}\approx 0.3$ for a black hole spinning at the Thorne limit ($a=0.998$).  Thus, even if the true AGN population had equal numbers of non-spinning and rapidly spinning objects, the rapid spinners would be over-represented in a flux limited survey by a factor of $(\eta_{\rm max}/\eta_0)^{3/2}\approx 11$ (assuming a Euclidean Universe).

Accounting for this selection effect, the dominance of rapidly rotating black holes is not at all surprising (Brenneman et al. 2011; Vasudevan et al. 2015).  Figure~3 (right panel) illustrates this  quantitatively by showing a histogram of the current spin measurements, overlaid with the distributions expected in a flux limited sample when the number of objects in the underlying population per unit dimensionless spin is $n(a)\propto a^p$, where $p=0,1,2$.  Repeating the caveat that this is not rigorously a flux-limited sample, the complete lack of objects below $a=0.4$ tends to argue against a flat distribution, $p=0$, across the full range of prograde spins $a\in(0,1)$.   However, the data are consistent with a flat distribution truncated at the low-end by $a\sim 0.3-0.4$.  Clearly, the selection effect resulting from the spin-dependent radiative efficiency is an important selection effect that needs to be fully accounted for in any attempts to extract population statistics from spin data.

\section{Fast winds and AGN feedback}

\begin{figure*}[t]
\hbox{
\psfig{figure=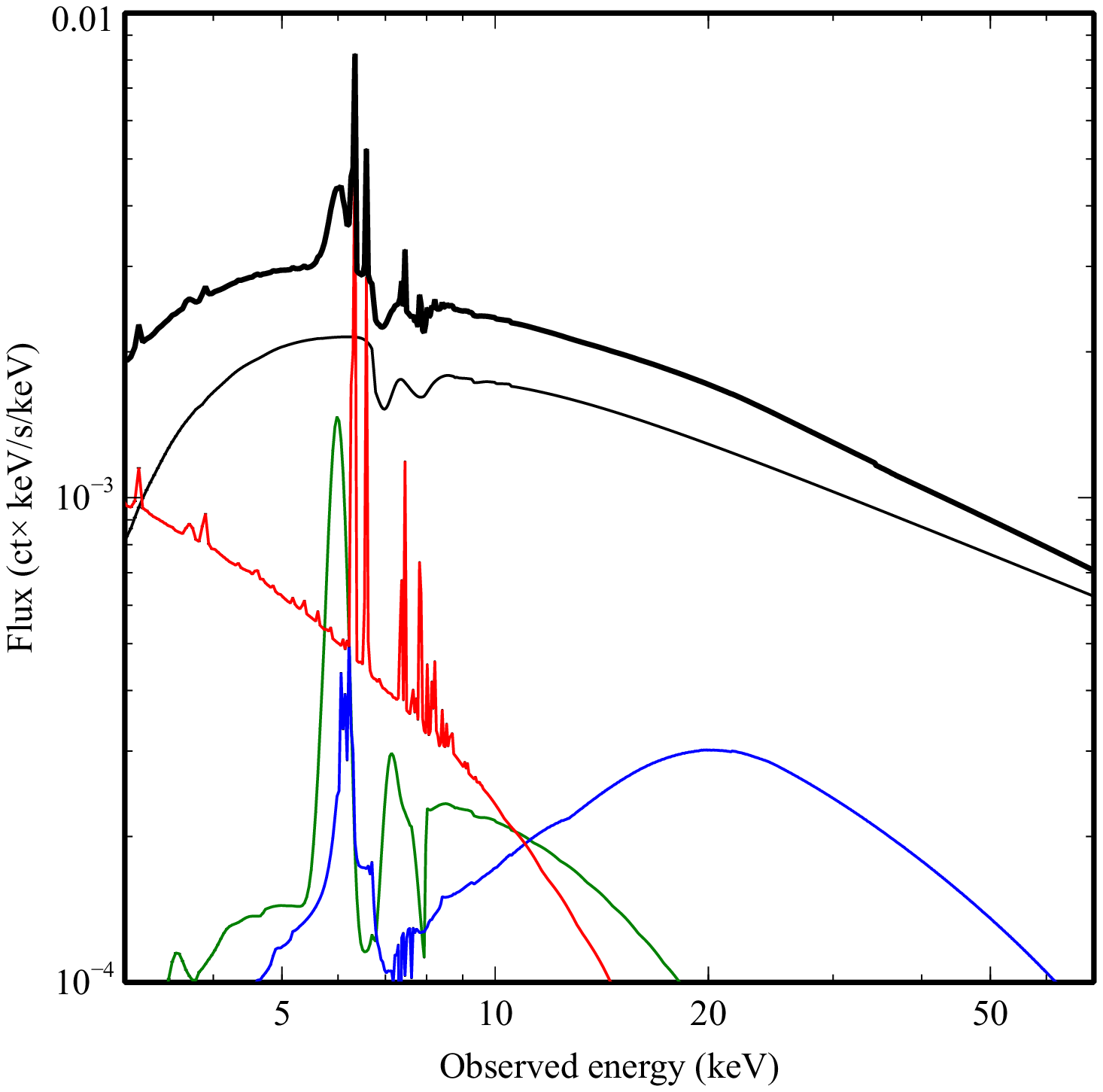,width=0.48\textwidth}
\hspace{0.3cm}
\psfig{figure=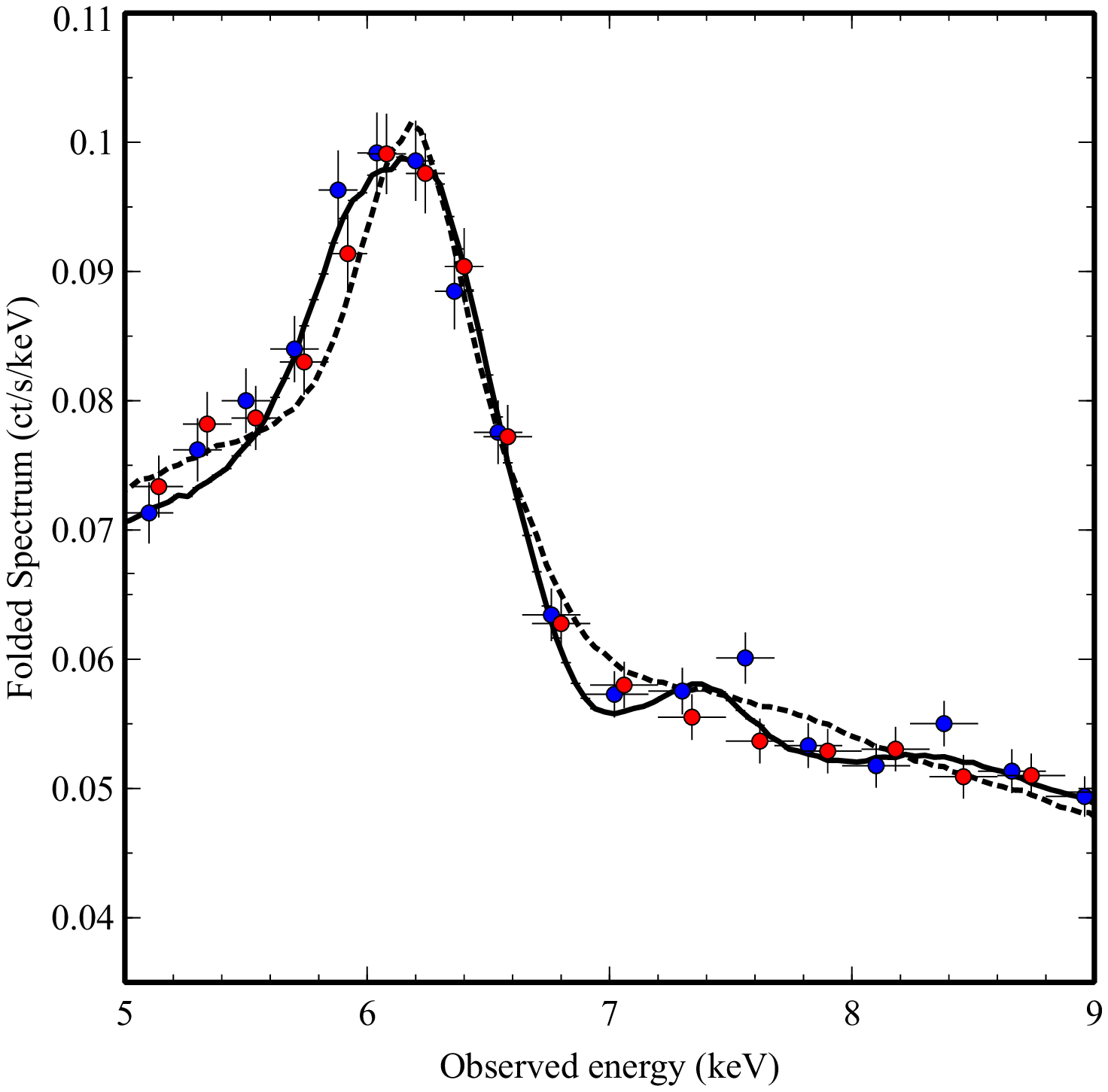,width=0.48\textwidth}
}
\caption{The NuSTAR view of the powerful radio galaxy Cygnus~A.  {\it Left Panel : }Spectral model resulting from fitting the 3--70\,keV NuSTAR spectrum.  The primary AGN continuum is affected by both neutral absorption and FeXXV/FeXXVI line absorption from a fast ionized wind (resulting in thin black line).  Emission from this wind, producing a P-Cygni type profile, is also seen (green line).  In addition, the NuSTAR data require an unbroadened but ionized reflection component (blue line).  Finally, the model includes the thermal emission of the surrounding cluster ICM (red line).  The sum of all model components give the observed spectrum (thick black line).   {\it Right Panel : }Folded spectrum from NuSTAR, focusing on the iron K-band; data from the two Focal Plane Modules are shown separately (FPMA=red, FPMB=blue).   Also shown are the (folded) best-fitting models with (solid line) and without (dotted line) the absorption/emission associated with the fast ionized wind. Figures reproduced from Reynolds et al. (2015).}
\end{figure*}

X-rays are an excellent probe of accretion disk winds.  These disk winds have attracted a substantial amount of attention since they are likely the agent by which luminous AGN can feedback on their host galaxies and regulate their own growth (thereby resulting in the $M-\sigma$ relation; Gebhardt et al. 2000, Ferrarese \& Merritt 2000).  Of particular relevance are the fast, highly-ionized, high-column density winds seen via the blue-shifted X-ray K-shell absorption lines of FeXXV and FeXXVI (Chartas et al. 2002, 2009; Pounds et al. 2003; Tombesi et al. 2010, 2013; Gofford et al. 2011).  Due to their high-velocities (often in the range $v\sim 0.1-0.3c$) these winds carry a substantial kinetic power which can exercise significant feedback on the larger scale galactic environment.  

\begin{figure*}[t]
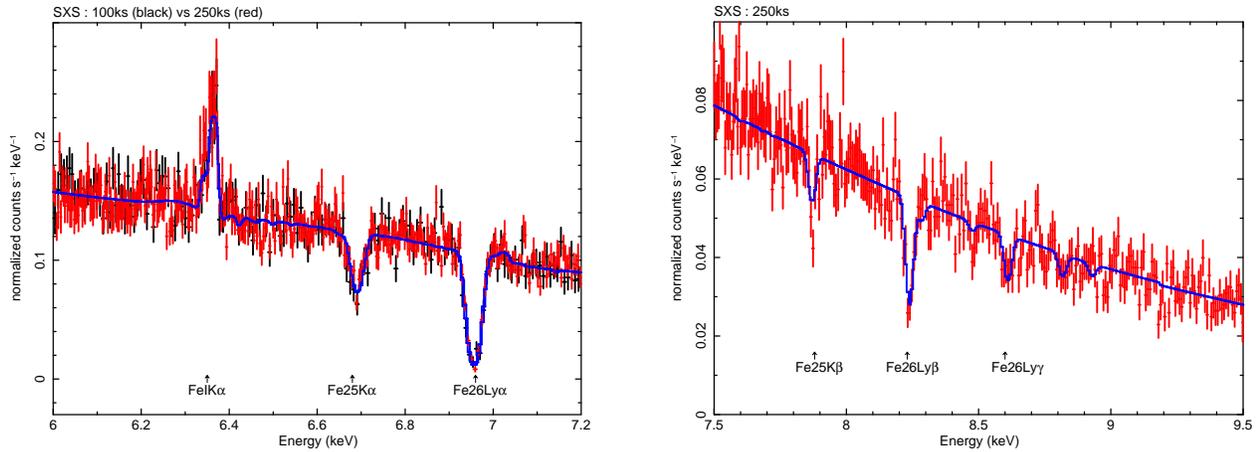

\hbox{
\psfig{figure=mcg6plot_ironband.ps,width=0.35\textwidth,angle=270}
\hspace{0.3cm}
\psfig{figure=mcg6plot_highironband_250ks.ps,width=0.35\textwidth,angle=270}
}
\caption{Simulated 250\,ks exposure of the Seyfert galaxy MCG--6-30-15 ($z=0.008$) with the Astro-H Soft X-ray Spectrometer (SXS), focusing in on the iron K$\alpha$ band (left panel) and higher order K-shell transitions (right panel). The Chiang \& Fabian (2011) model was used for the spectral simulation, and includes reflection from distance/slow material producing the narrow iron fluorescence line at 6.4\,keV (rest-frame) as well as absorption from a highly ionized (ionization parameter $\log\xi=3.8$) high column-density ($N_H=2\times 10^{23}\,{\rm cm}^{-2}$) wind outflowing from the AGN at $2000\,{\rm km}\,{\rm s}^{-1}$.  }
\end{figure*}

Given their relevance for galaxy evolution, it is interesting to study these winds in luminous (high Eddington ratio) nearby objects which may act as prototypes for AGN feedback processes that were common at higher redshift.  Recent process has been dramatic. NuSTAR has revealed that the strong blue-shifted FeXXVI absorption line in the luminous radio-quiet quasar PDS456, previously discovered by XMM-Newton (Reeves et al. 2003), has an associated emission component producing a P-Cygni profile (Nardini et al. 2015).  For the first time, this allows the determination of the solid angle subtended by this $v\sim 0.25c$ wind (as seen from the center of the AGN), and removes one of the big uncertainties that existed in estimates of the wind's kinetic power.  Nardini et al. (2015) estimate a kinetic power of $L_{\rm kin}\sim 2\times 10^{46}\,{\rm erg}\,{\rm s}^{-1}$ in PDS456, sufficient to unbind the galactic bulge in $10^7$\,yr (if the wind's energy were to be captured with 100\% efficiency).   Using Suzaku, Tombesi et al. (2015) found a similarly fast and ionized wind in the ultra-luminous infrared galaxy IRAS~F1111913257.   They showed that the extended molecular (OH) outflow in this object revealed by Herschel can be driven by this fast AGN wind, provided that the driving is in an energy-conserving mode (Faucher-Gigu\'ere \& Quataert 2012; King \& Pounds 2003).  This was the first time that the fast ionized AGN wind and a galactic-scale molecular outflow could be related in a single object.  

The form of feedback mediated by these winds is often referred to as quasar-mode feedback.  The other established mode is jet-mediated, or radio-mode, feedback that operates in cool core galaxy clusters and giant elliptical galaxies.  The notion that these two modes of feedback are distinct and mutually exclusive has recently been challenged by NuSTAR observations of Cygnus~A, the low-$z$ prototypical powerful radio galaxy whose jets are clearly exercising feedback on the intracluster medium of its host cool core galaxy cluster.  By allowing the clearest study yet of this absorbed AGN, NuSTAR reveals a fast ($v\sim 0.05-0.1c$), highly-ionized, high-column ($N_H>3\times 10^{23}\,{\rm cm}^{-2}$) density wind that carries a kinetic energy of at least $L_{\rm kin}\sim 2\times 10^{45}\,{\rm erg}\,{\rm s}^{-1}$ (Reynolds et al. 2015; see Fig.~4).  Re-emission from the wind, producing a P-Cygni profile, was also seen in this case, again demonstrating that this is a wide angle wind subtending a significant fraction of the sky as seen from the central X-ray source.   The case of Cygnus~A demonstrates that wind-mediated quasar-mode feedback and jet-mediated radio-feedback can occur simultaneously. 

The future prospects for studying these highly ionized winds is extremely bright due to the imminent deployment of the Soft X-ray Spectrometer (SXS) on the Astro-H mission (see Astro-H White Paper ``AGN Winds''by Kaastra et al. 2014).  As an illustration, Fig.~5 shows the iron-K band from a 250\,ks simulated Astro-H/SXS observation of the bright Seyfert galaxy MCG--6-30-15.  This is known to possess a highly ionized wind outflowing at $\sim 2000\,{\rm km}\,{\rm s}^{-1}$ from previous Chandra High-Energy Transmission Grating (HETG) studies (Young et al. 2005; Chiang \& Fabian 2011), but the detailed properties of that wind could not be studied.  The SXS will permit very high signal-to-noise detections of the FeXXV-K$\alpha$ and FeXXVI-K$\alpha$ lines (Fig.~5 left panel), giving a detailed view of the kinematics of the absorber. But, it will also permit the detection of higher order lines (possibly up to FeXXVI-K$\delta$; Fig.~5 right panel), allowing for saturation and partial covering effects to be fully disentangled in our determination of the ionization state and column density of the wind.  Furthermore, it is likely that the known absorber, i.e. that seen by Chandra/HETG and included in this simulation, is just the tip of the iceberg.  The discovery space for uncovering previously unknown wind components, some of which may be dominant terms in the feedback budget if they are sufficiently fast, is very large.  

\section{Outlook}

This paper has given a brief survey of some of the most interesting developments in AGN physics that have been enabled by X-ray spectroscopy.  The future of this field is very bright, especially given the imminent launch of Astro-H.  We have already discussed how the broad-band pass enabled by the HXI and SGD on Astro-H will impact our understanding of X-ray emitting coronae, as well as the opportunity for studying fast and ionized winds (which most likely drive quasar-mode feedback) with the superior spectral resolution of the Astro-H SXS.  Astro-H will also probe the nature of obscuring matter in Seyfert 2s and Compton-Thick AGN, using the profile of the iron fluorescence line to determine the kinematics and hence elocution of the matter, and the shape of the Compton shoulder to determine the geometry of the obscuring matter distribution.  

Looking a little further ahead, the prospects of X-ray polarimetry will add an entirely new dimension to these studies.  For example, the Polarimeter for Relativistic Astrophysical X-ray Sources (PRAXyS), currently in Phase-A study for NASA's current Small Explorer round, will permit spectropolarimetry of a bright AGN across the 2--10\,keV band.  Such data will provide a powerful new tool for studying X-ray reflection from the accretion disk or, indeed, scattering from any non-spherical distribution within the AGN central engine.  With two X-ray polarization missions currently under study in the United States, as well as the X-ray Imaging Polarimetry Explorer (XIPE) under study for the ESA's M4 slot, the chances of realizing the promise of X-ray spectropolarimetry in the next decade is extremely high. 

\smallskip

The author thanks Laura Brenneman, Andy Fabian, Anne Lohfink, and Francesco Tombesi for insightful discussions that have informed much of the discussion in this paper.   The author is grateful for financial support from the Simons Foundation (through a Simons Fellowship in Theoretical Physics), a Sackler Fellowship (hosted by the Institute of Astronomy, Cambridge), and NASA under grant NNX14AF86G.


\appendix

\end{document}